\documentclass[]{spie}  

\usepackage{amsmath,amsfonts,amssymb}
\usepackage{graphicx}
\usepackage[colorlinks=true, allcolors=blue]{hyperref}
\usepackage{caption}
\usepackage{subcaption}
\usepackage{amsmath}

\title{Characterization of avalanche photodiodes (APDs) for the CUbesat Solar Polarimeter (CUSP) mission}

\author[a]{Cologgi~F.}
\author[a]{Alimenti~A.}
\author[b]{Fabiani~S.}
\author[a]{Torokthii~K.}
\author[a]{Silva~E.}
\author[b]{Del~Monte~E.}
\author[f]{Baffo~I.}
\author[d]{Bonomo~S.}
\author[c]{Brienza~D.}
\author[h]{Campana~R.}
\author[i]{Centrone~M.}
\author[g]{Contini~G.}
\author[b]{Costa~E.}
\author[c]{Curatolo~A.}
\author[d]{Cucinella~G.}
\author[b]{DevAngelis~N.}
\author[h]{De~Cesare~G.}
\author[g]{Del~Re~A.}
\author[b]{Di~Cosimo~S.}
\author[d]{Di~Filippo~S.}
\author[b]{Di~Marco~A.}
\author[b]{Di~Persio~G.}
\author[c]{Donnarumma~I.}
\author[f]{Fanelli~P.}
\author[g]{Leonetti~P.}
\author[e]{Locarini~A.}
\author[b]{Loffredo~P.}
\author[b]{Lombardi~G.}
\author[j]{Minervini~G.}
\author[e]{Modenini~D.}
\author[b]{Muleri~F.}
\author[c]{Natalucci~S.}
\author[d]{Nigri~A.}
\author[d]{Perelli~M.}
\author[b]{Rossi~M.}
\author[b]{Rubini~A.}
\author[b]{Scalise~E.}
\author[b]{Soffitta~P.}
\author[c]{Terracciano~C.}
\author[e]{Tortora~P.}
\author[c]{Zaccagnino~E.}
\author[g]{Zambardi~A.}

\affil[a]{Dept. of Industrial, Electronic and Mechanical Engineering, Roma Tre University, Via V. Volterra 62, 00146 Rome, Italy}
\affil[b]{INAF-IAPS, via del Fosso del Cavaliere 100, 00133, Rome, Italy}
\affil[c]{Agenzia Spaziale Italiana, via del Politecnico snc, 00133 Rome, Italy}
\affil[d]{IMT s.r.l., via Carlo Bartolomeo Piazza 30, 00161 Rome, Italy}
\affil[e]{Dept. of Industrial Engineering and Interdepartmental Center for Industrial Aerospace Research, Alma Mater Studiorum Università di Bologna, Via Fontanelle 40, 47121 Forl\`i, Italy}
\affil[f]{DEIM, Università degli studi della Tuscia, Largo dell'Università, 01100 Viterbo, Italy}
\affil[g]{SCAI Connect s.r.l., Via Vincenzo Lamaro 51, 00173 Roma, Italy}
\affil[h]{Dip. di Ingegneria dell'Impresa ``Mario Lucenti'', Università degli Studi di Roma ``Tor Vergata'', Via Cracovia 50, 00133 Roma, Italy
}
\affil[i]{INAF-OAR, Via Frascati 33, 00040, Monte Porzio Catone, Italy}
\affil[j]{INAF Headquarters, Viale del Parco Mellini 84, 00136, Roma, Italy}

\authorinfo{Send correspondence to F.S., E-mail: sergio.fabiani@inaf.it}

\begin{document} 
\maketitle

\begin{abstract}
The CUbesat Solar Polarimeter (CUSP) project is a CubeSat mission orbiting the Earth aimed to measure the linear polarization of solar flares in the hard X-ray band by means of a Compton scattering polarimeter. CUSP will allow the study of the magnetic reconnection and particle acceleration in the flaring magnetic structures of our star. CUSP is a project in the framework of the Alcor Program of the Italian Space Agency aimed at developing new CubeSat missions. It is approved for a Phase B study. 
In this work, we report on the characterization of the Avalanche Photodiodes (APDs) that will be used as readout sensors of the absorption stage of the Compton polarimeter. We assessed the APDs gain and energy resolution as a function of temperature by irradiating the sensor with a \textsuperscript{55}Fe radioactive source. Moreover, the APDs were also characterized as being coupled to a GAGG scintillator.

\end{abstract}


\section{INTRODUCTION}
\label{sec:intro}  
The CUbesat Solar Polarimiter (CUSP) project aims to measure the polarization of the X-ray emissions from solar flares. This will open new possibilities in the study of these highly energetic events \cite{tandberg1988physics}, both for an in-depth comprehension of the phenomenon and to develop solutions to protect human technological infrastructures from the potential risks related to solar flares \cite{joselyn1992impact}. 

The payload of CUSP is made up of an X-ray Compton polarimeter composed of plastic and inorganics scintillators \cite{fabiani2018instrumentation,fabiani2022cusp, fabiani2024cusp}. The light produced by the scattering and absorption of the X-ray photons with the scintillators is read by an array of avalanche photodiodes (APS) which will work in time coincidence with the output signal produced by multianode photomultiplier tubes (MAPMT). An accurate characterization and calibration of the sensitive devices is fundamental for the polarimeter performances. In this work, we focus on the characterization of the APD \cite{ikagawa2003performance}. The APD gain $G_{APD}$ and energy resolution $\delta E/E$ are measured as a function of the bias voltage $V_b$, and temperature $T$, while a \textsuperscript{55}Fe radioactive sample is used as an X-ray photon source and placed first in direct contact with the APD and then interposing between these the Gd$_3$Al$_2$Ga$_3$O$_{12}$ (GAGG) inorganic scintillator that will be used in the final configuration of the polarimeter. $G_{APD}(V_{bias},T)$ is then measured in the temperature range $-20<T/(^\circ\mathrm{C})<60$ between 260~V and 410~V. In particular, the measurement of $G_{APD}(V_{bias},T)$ will be fundamental to designing the feedback loop control system able to keep $G_{APD}$ stable, during the orbiting of the CubeSat, just by acting on $V_{bias}$.

In this work, we describe the measurement system designed to perform this calibration, and the first results obtained in a climate chamber on two APDs are shown.

\section{The Measurement system }
\label{sec:MeasSyst}  

\begin{figure}
    \centering
    \includegraphics[width=0.9\textwidth]{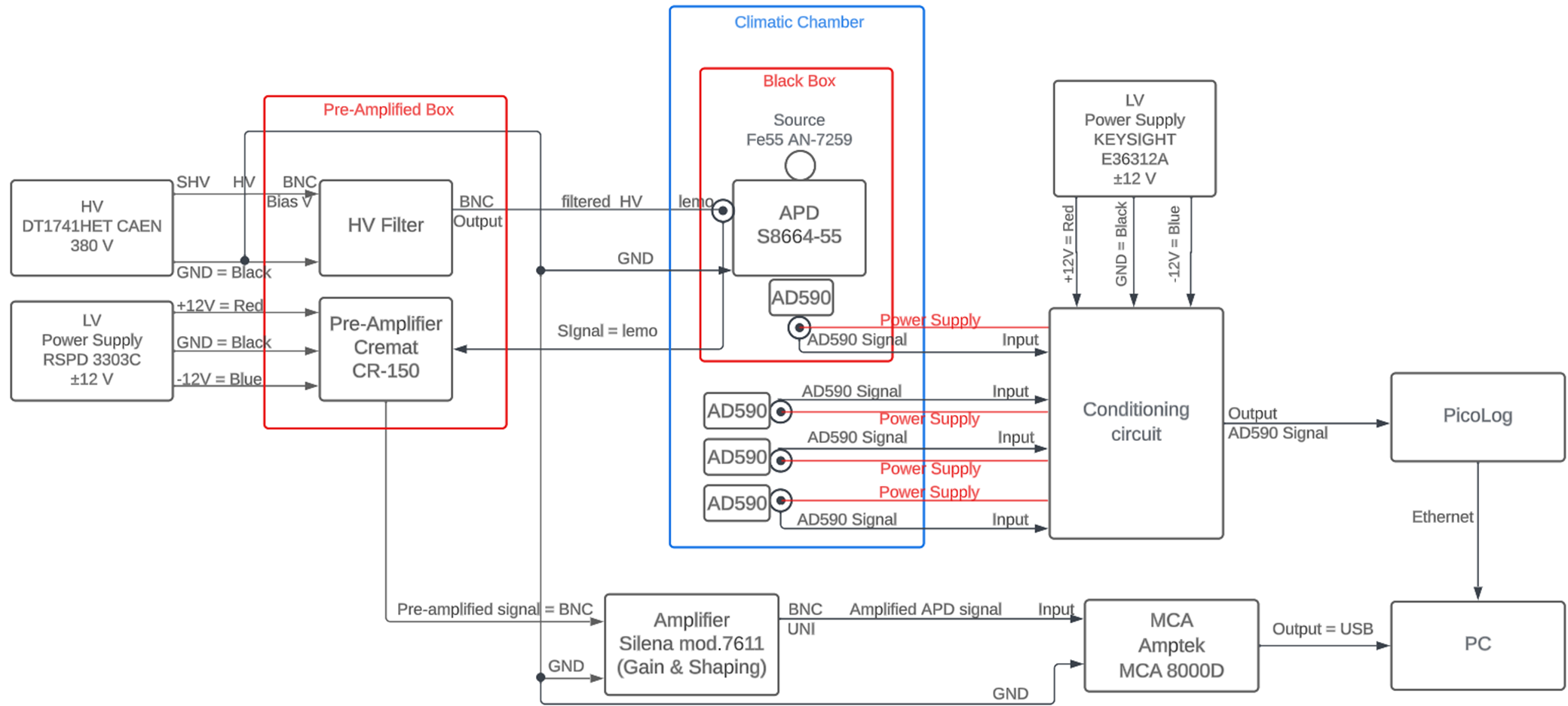}
    \caption{Block diagram of the set-up used for the characterization of the APD.}
    \label{fig:System}
\end{figure}

The block diagram of the designed measurement system is shown in \figurename~\ref{fig:System}.
The APD bias voltage $V_{bias}$ is generated by the High Voltage (HV) source DT1471HET CAEN and filtered with the HV filter on the Cremat CR-150 evaluation board. Once the APD is illuminated by a radioactive source, the charge pulses are pre-amplified by the Cremat CR-110R2.2 charge amplifier, setting a 1.4~V/pC conversion factor, and a characteristic time $\tau\approx1.4\,\mathrm{\mu s}$. Then, in order to reduce the pulse pileup, the Silena 7611 spectroscopy amplifier is used for the pulse shaping. Finally, the Amptek MCA8000D multi-channel analyzer is used in the pulse-height analysis mode to record the number of events detected on each channel. To control $T$, the APD is set in a climate chamber (Angelantoni, model DY2000), and four AD590 $T$ sensors are placed in the chamber for fine-tuning and measurement of $T$. 

The APD gain $G_{APD}$ measurements are performed by acquiring the amplitude of the voltage pulses in input to the MCA. Thus, the conversion factor between the charge generated by each impinging photon and the amplitude of the voltage pulses is fundamental for the accurate $G_{APD}$ measurement. However, the evaluation of this conversion factor would require an accurate knowledge of all the amplification/attenuation introduced by each component in the measurement system. An easier alternative consists in using a reference APD with known $G_{APD}$, and using this to calibrate the realized measurement system. In our case the Si S8664-55 Hamamatsu photodiode, serial number AA4400, was used as reference ($G_{ref}=50\pm1$ at $\sim25\,^\circ$C and with $V_{bias}\approx433$~V). Thus, by exploiting the MCA linearity, $G_{APD}$ is evaluated as follows:
\begin{equation}\label{eqn:Gcal}
    G_{APD}=\frac{Ch_{pk} G_{ref}}{Ch_{ref}}\;,
\end{equation}
where $Ch_{pk}$ represents the MCA channel number at which the pulses histogram peak is detected, and $Ch_{ref}=1052\pm1$ is the MCA channel corresponding to $G_{ref}$.

Finally, the energy resolution $\delta E/E$ is measured by fitting the histogram peaks with Gaussian curves and evaluating:
\begin{equation}
    \frac{\delta E}{E}=\frac{\mathrm{FWHM}}{Ch_{pk}}=2.35\frac{\sigma_{pk}}{Ch_{pk}}\;,
\end{equation}
where $\sigma_{pk}$ is the measured standard deviation, and $FWHM$ the corresponding full width at half maximum.

\section{Results and discussion}
\label{sec:Results}  
\begin{figure}
    \centering
    \includegraphics[width=0.5\textwidth]{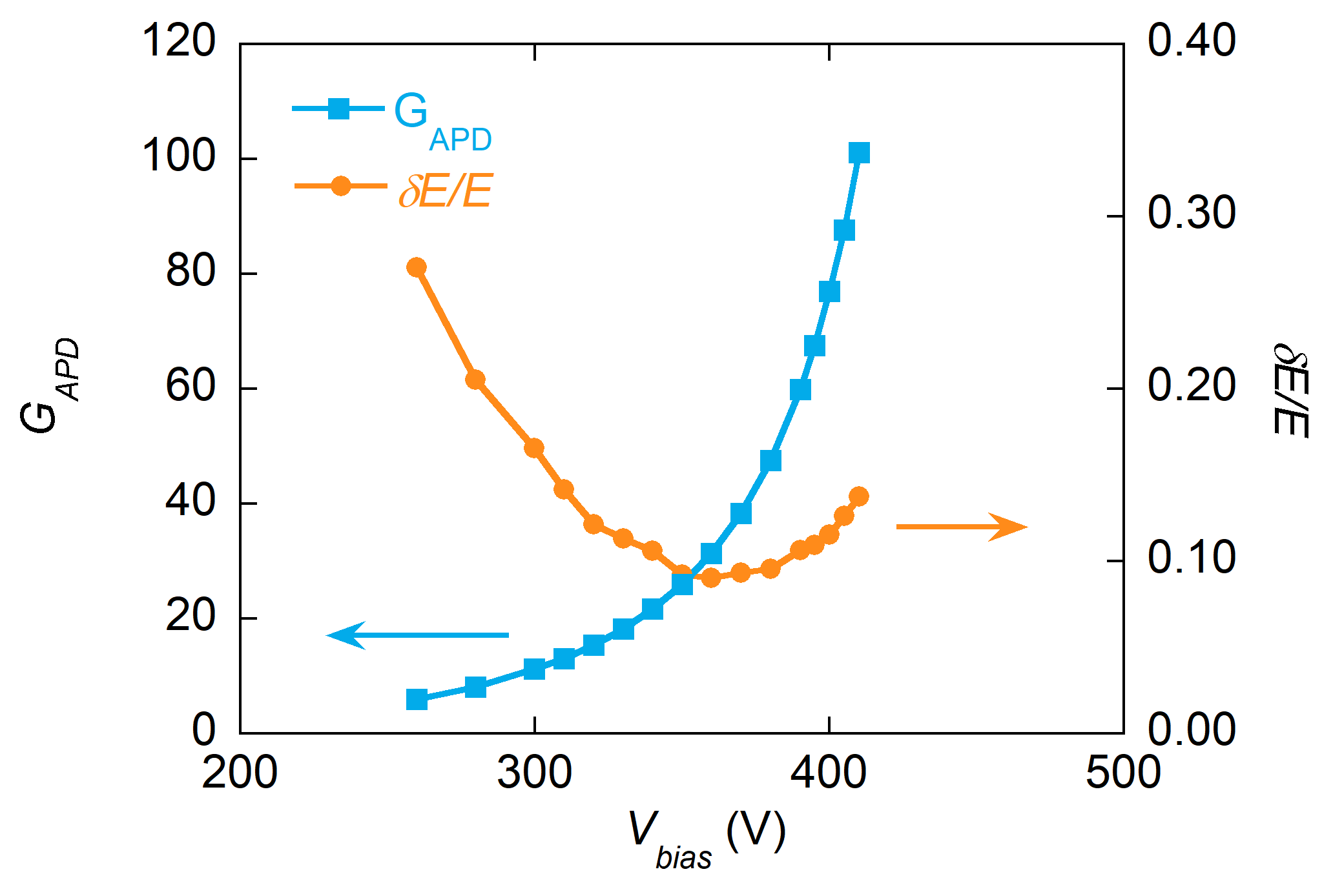}
    \caption{APD gain $G_{APD}$ (blue squares--left scale) and energy resolution $\delta E/E$ (orange circles -- right scale) as a function of the applied bias voltage $V_{bias}$. Data are acquired by placing the \textsuperscript{55}Fe samples directly illuminating the APD.}
    \label{fig:GEresvsVbias}
\end{figure}

First, the shaping time $t_s$ of the spectroscopy amplifier was set by reducing it below the threshold for which $\delta E/E$ was observed to no longer be sensitive to $t_s$. This happens for $t_s<3\;\mu$s. 
Once $t_s$ is set up, the $V_{bias}$ effect on $\delta E/E$ and $G_{APD}$ is first evaluated at room temperature. Thus a \textsuperscript{55}Fe sample is placed in contact with the reference APD and $V_{bias}$ changed. The result is shown in \figurename~\ref{fig:GEresvsVbias}. The measurements shown in \figurename~\ref{fig:GEresvsVbias} allow us to assess: (i) the calibration of the measurement system by exploiting Eq.\eqref{eqn:Gcal}, (ii) the optimum $V_{bias}$ useful to minimize $\delta E/E$.

\begin{figure}
\centering
\begin{subfigure}{.5\textwidth}
  \centering
  \includegraphics[width=1\linewidth]{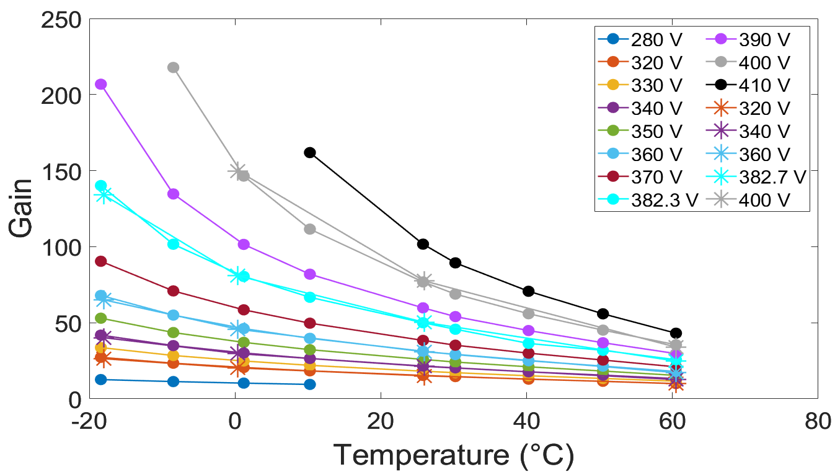}
  \caption{}
  \label{fig:sub1}
\end{subfigure}%
\begin{subfigure}{.5\textwidth}
  \centering
  \includegraphics[width=1\linewidth]{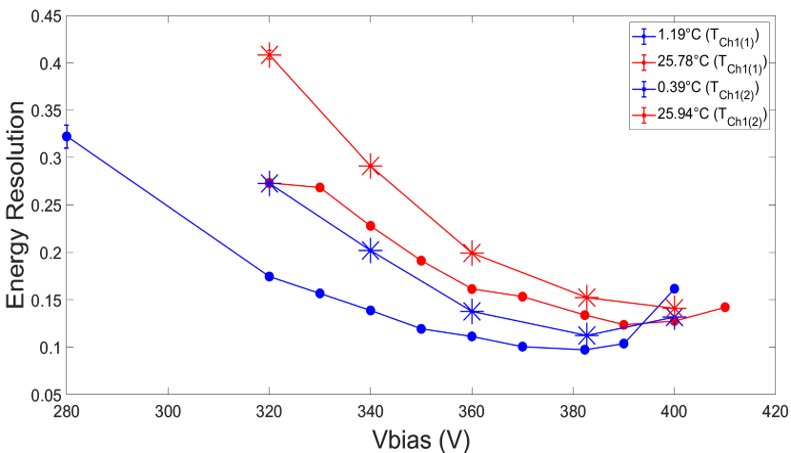}
  \caption{}
  \label{fig:sub2}
\end{subfigure}
\caption{(a) The gain $G_{APD}$ measurements as a function of temperature $T$ and bias voltage $V_{bias}$. (b) The energy resolution $\delta E/E$ measurements as a function of temperature $T$ and bias voltage $V_{bias}$. Both plots show a comparison between $G_{APD}$, and $\delta E/E$, when measured on two APDs (full circles and asterisks).}
\label{fig:measClimChamber}
\end{figure}

The so-calibrated system was then used to characterize the response of two APDs in a climate chamber. The temperature was swept in the range $-20<T/(^\circ\mathrm{C})<60$ in steps of $10\;^\circ\mathrm{C}$, while the APD bias voltage in the range $280<V_{bias}/(\mathrm{V})<410$ in steps of 10~V. The results are shown in \figurename~\ref{fig:measClimChamber}. $G_{APD}$ is shown to increase when $T$ is decreased, as expected and in agreement with the literature \cite{ikagawa2003performance}. The measurement of the surface $G_{APD}(T,V_{bias})$ thus allows tuning a feedback loop useful to hold $G_{APD}$ stable, by acting on $V_{bias}$, despite of the unavoidable $T$ variation of the polarimeter in the orbiting CubeSat. This, however, will cause variations of $\delta E/E$ which in turn is dependent on $T$ and $V_{bias}$ as experimentally found and shown in \figurename~\ref{fig:measClimChamber}-(b).

Finally, the results show a small variability of the main features of the obtained calibration curves among different APDs. In future, the work will be extended to a larger number of APDs to perform statistical analysis. 

\acknowledgments 
 Activity funded by ASI-INAF CUSP phase A contract 2022-4-R.0.

\bibliography{main} 
\bibliographystyle{spiebib} 

\end{document}